\def\gr{$\gamma$-ray }

\documentstyle[12pt,aaspp4]{article}

\begin{document}

\title{The Pulsar Contribution to the Diffuse Galactic $\gamma$-ray Emission}
\author{Martin Pohl\altaffilmark{1} and Gottfried Kanbach}
\affil{MPE, Postfach 1603, 85740 Garching,Germany}
\altaffiltext{1}{Current address: Danish Space Research Institute, Juliane Maries Vej 30,
2100 K\o benhavn \O ,Denmark}
\author{Stan D. Hunter}
\affil{GSFC, Code 662, Greenbelt, MD 20771, USA}
\and
\author{Brian B. Jones}
\affil{W.W. Hansen Laboratory, Department of Physics, Stanford University,
Stanford, CA 94305}
\authoremail{mkp@mpe-garching.mpg.de}

\begin{abstract}
There is active interest in the extent to which
unresolved \gr pulsars contribute to the galactic diffuse emission, and
whether unresolved \gr pulsars could be
responsible for the excess of diffuse galactic emission above
1 GeV which has been observed by EGRET.
The diffuse \gr intensity due to unresolved pulsars is directly linked
to the number of objects which should be observed in the
EGRET data. We can therefore use our knowledge of the unidentified EGRET sources
to constrain model parameters like the pulsar birth rate, their beaming angle, etc.
This analysis is based only on the properties of the six pulsars which have
been identified in the EGRET data, and is independent of choice of a pulsar
emission model.

We find that pulsars contribute very little to the diffuse emission at lower
energies, whereas above 1 GeV they can account for 18\%
of the observed intensity in selected regions for a reasonable number of 
directly observable \gr pulsars ($\sim$14). The latitude distribution
of the diffuse emission caused by unresolved pulsars is narrower than 
that of the observed diffuse emission.
While the excess above 1 GeV \gr energy
is observed up to at least 6$\arcdeg$-8$\arcdeg$ degrees off the plane, 
the pulsar contribution would be small there. Thus pulsars do
significantly contribute to the diffuse galactic $\gamma$-ray emission
above 1 GeV, but they can not be made responsible for all the discrepancy
between observed intensity and model predictions in this energy range.
\end{abstract}

\keywords{Gamma rays: observations; pulsars: general}

\section{Introduction}

A recent analysis of the diffuse galactic \gr emission in the energy range
of 30 MeV to 30 GeV (\cite{hunt97}) indicated
that the spatial structure and total intensity of the emission observed
by EGRET can be well understood as the result of interactions between
cosmic rays and the interstellar medium in addition to an isotropic
extragalactic background. The main emission mechanisms are $\pi^0$ 
production and decay in inelastic collisions between cosmic ray nucleons and
thermal gas; bremsstrahlung of relativistic electrons in thermal gas; and
inverse-Compton scattering of ambient mm-wave, infrared, and optical
photons by highly relativistic electrons. The limitation of these studies
is the inability to separate truly diffuse \gr emission from that of
unresolved discrete sources.

At energies above 1 GeV the models predict only roughly 60\%
of the observed intensity. This effect is most significant in the 
galactic plane, but is not restricted to it.
This discrepancy might be explained in two different ways.
First, it may be that the Galactic cosmic ray protons and/or electrons have a harder spectrum beyond 10 GeV than those observed in the solar vicinity.
Second, there may be a
number of unresolved point sources in the galactic plane with appropriate
spectra.
The \gr spectrum of these sources would have to be harder than that
of the galactic diffuse emission below a few GeV and may roll over beyond
a few GeV. The only known objects with such spectra are pulsars,
especially Geminga-type pulsars.

The attempt of this paper is twofold. We want to provide constraints on
the general contribution of the most likely input from
discrete sources -- pulsars --
to the diffuse galactic \gr emission. We also want to find out whether or not
unresolved pulsars can account for the observed excess at high \gr energies.
Previous work has addressed the former problem (\cite{bk92},
\cite{yadi95}, \cite{sd96}) primarily on the basis of pulsar models, for
which the input parameters are not always well known.
Conversely, \gr data have been used to limit the emission from millisecond
pulsars (\cite{bmp97}). Here we use a different strategy: instead of
using models we will base our analysis solely on the properties of the 
six pulsars observed by EGRET. Clearly, the main limitation of this approach
is that the six objects are not necessarily representative
for the whole population.
The great advantage of our method however is that all spectral information
can be taken into account. This will allow us to address 
whether pulsars can account for 
the observed excess at high \gr energies.

In general, pulsars are less luminous when older and they often 
 exhibit spectral cut-offs at high \gr energies. Since the
luminosity and spectral evolution 
provide the best parameters from which to estimate
the contribution of pulsars to the diffuse \gr emission,
we will derive the luminosity of an `ideal' pulsar as a function of
its age and its observed energy directly from the EGRET data.
Since this method will use the observed intensity of the pulsars
at different energies, it will automatically account for the hard
spectrum and for possible cut-offs without using power-law fit results
on individual data.
We will simultaneously calculate the diffuse intensity
due to pulsars and the numbers of directly identifiable objects. The latter
provides a constraint on otherwise weakly determined parameters like the
birth rate and the beaming angle of pulsars, since
the diffuse \gr intensity due to unresolved pulsars scales almost
linearily with the number of resolved pulsars.

\section{The model}
\subsection{The intensity-age distribution of pulsars}
EGRET has identified six pulsars by their light curves. The
differential \gr photon spectra of these sources have been measured by
\cite{fier95} from pulsed analysis. The restriction to pulsed analysis
neglects part of possible unpulsed emission but allows us to obtain 
better spectra of the weak pulsars. The EGRET sensitivity is rather poor
at the lowest energies so that we will concentrate on nine energy bands
spanning 50 MeV to 10 GeV, for which we have the observed photon flux
$S_i (E_k)$ of pulsar $i$ in energy band $E_k$. 
In an analogous way to the calculation for absolute brightnesses in optical astronomy, 
we multiply these
fluxes by the square of the pulsar's distance $D_i$ 
to find fluxes we would observe from these sources if they were at unit
distance.
So in each of the nine energy bins we get the `absolute' \gr
intensity and the age of the six pulsars from our data. The basis
of our modelling is the assumption that pulsars 
exhibit a correlation between
absolute intensity and age. The slope and normalisation of this correlation
may be different at different \gr energies. This correlation,
which describes sort of an ideal pulsar, will then be used to calculate
the diffuse emission from unresolved pulsars together with the number
of directly observable objects.

There is no physical necessity why such a correlation should exist. The
best evidence comes from the data themselves: the $\chi^2$ sums we obtain
indicate that a correlation is an appropriate description of the
data. In reality one should expect such a correlation to be noisy, i.e.
that there is a probability distribution with a certain dispersion around
the mean of the correlation, but in our case the limited number of degrees
of freedom forces us to use the most simple correlation model, a
power-law:

\begin{equation}
S_m(t,E_k) = 10^{y_k}\, \left({{t}\over {10^4\ {\rm years}}}\right)^{b_k}
\quad {\rm kpc^2/ cm^2 / sec / MeV}
\end{equation}
This relation
can be fitted to the data
on the basis of weighted least squares for all energy bins.
This is equivalent to assuming Gaussian errors in the measured luminosities.
The kpc$^2$ part of the
units comes from our choice for the distance normalization. 
The distance $D_i$ and its uncertainty can be generally derived
on the basis of the pulsar dispersion
and rotation measure (\cite{tayl93}) and only for extremely
nearby objects like Geminga by parallax measurements (\cite{cara96}).
The age of pulsars can be estimated from the ratio of period and period
derivative. The uncertainty of this measure of age is large, especially since
pulsars often do not slow down solely by dipole radiation. We will use a 
factor of two for the age uncertainty $\delta t_i$, except for the
Crab for which the true age is taken with a nominal uncertainty of 10\%.

Three sources of uncertainty have to be considered in the fit: uncertainties
in intensity, in distance, and in age. While the intensity error can be 
assumed to follow a Gaussian probability distribution, both the errors
in age and distance appear in powers, so that their effective probability
distributions are definitely not Gaussian.
 Furthermore, the way
ages and distances are measured leads us to believe that a Gaussian
probability function is not a fair description of the actual error
distributions.
The errors in these parameters are sometimes of the same order as the estimates,
clearly implying that their distributions are asymmetric.
We can account for these effects and still use the $\chi^2$ method,
if we assume that the error distributions for the logarithms of age and
distance are Gaussians.

To account for the different uncertainty distributions in the three paramters,
we have separated the $\chi^2$-summation
into three components.
The total $\chi^2$ (\ref{totalchi2}) is found by combining the contributions of the 
intensity (\ref{schi2}), the distance (\ref{dchi2}), and the age (\ref{tchi2}).
\begin{equation}
\label{totalchi2}
\chi_i^2 = {{1}\over {\chi_{i,1}^{-2}+\chi_{i,2}^{-2}+\chi_{i,3}^{-2}}}
\end{equation}
where

\begin{equation}
\label{schi2}
\chi_{i,1} = {{S_m(t,E_k) -S_i (E_k)\,D_i^2}\over {D_i^2\,\delta S_i (E_k)}}
\end{equation}

\begin{equation}
\label{dchi2}
\chi_{i,2} = {{\log S_m(t,E_k) -\log S_i (E_k) - 2 \log D_i}\over
{2\delta \log D_i}}
\end{equation}

\begin{equation}
\label{tchi2}
\chi_{i,3} = {{\log S_m(t,E_k) -\log S_i (E_k) - 2 \log D_i}\over
{b_k \delta \log t}}
\end{equation}

\begin{figure}
\plotone{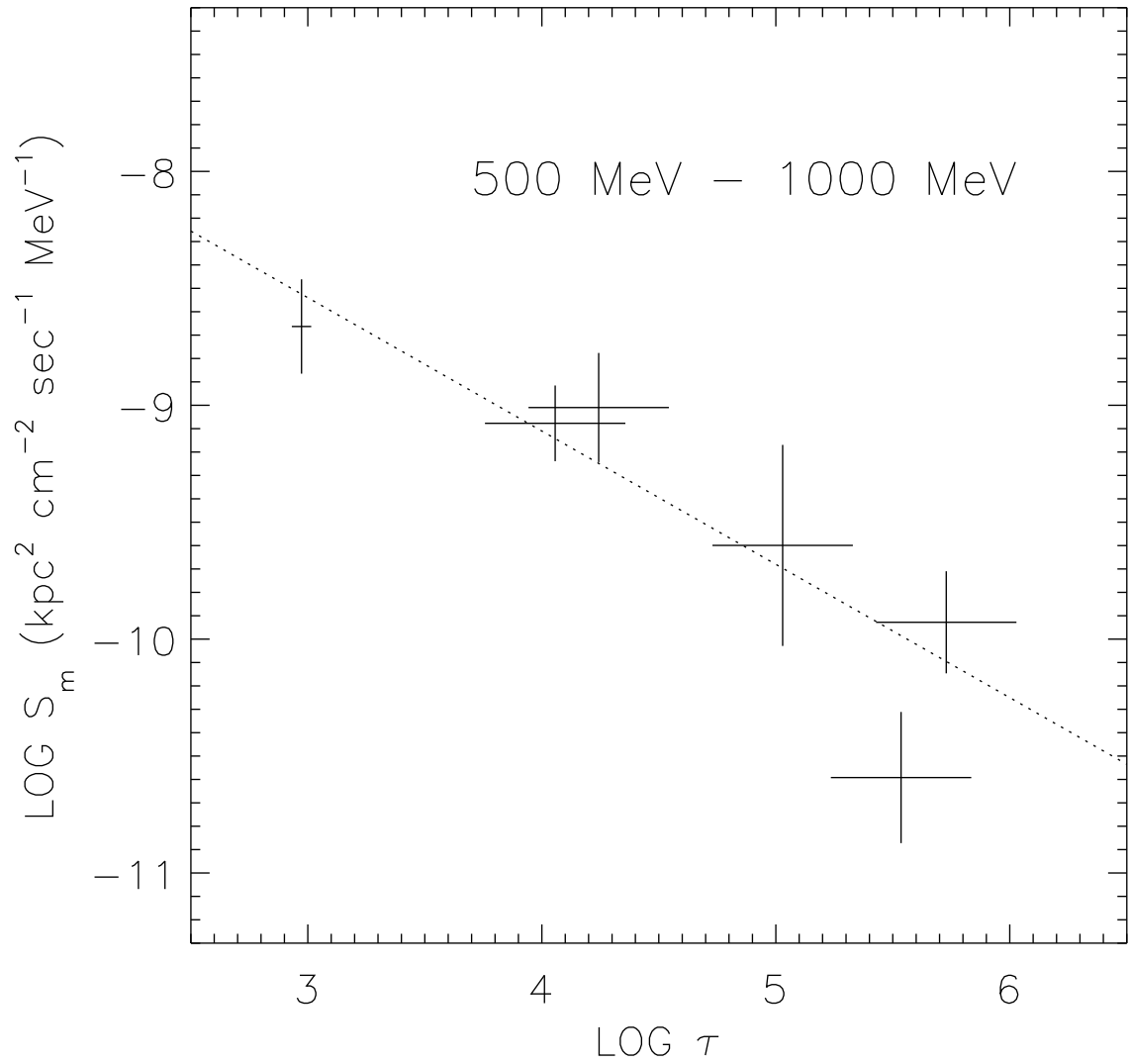}
\caption{This plot shows for a specific energy range the model
fit to the data of the six observed pulsars normalized to a distance
of 1 kpc. \label{lum}}
\end{figure}

\placefigure{lum}

An example of how a power-law relation fits to the pulsar data after
distance normalization is shown in Fig. \ref{lum} where all sources of uncertainty
are included as vertical and horizontal error bars.
The results of the fitting procedure are shown in Table \ref{tbl-1}, where the
1$\sigma$ uncertainty of the fit parameters is estimated from the usual
$\chi_{min}^2 +1$ condition.

\placetable{tbl-1}

\subsection{The fraction of unresolved pulsars}

For each source we can define a critical distance up to which the source
can be detected directly, and beyond which it would contribute to the
diffuse emission. The true \gr flux threshold of the EGRET data has
a rather awkward distribution on the sky. This is partly caused by the
structure of the galactic background emission and partly by the uneven
exposure. In addition to this, the broad point spread function of EGRET
causes point sources to enhance the background over regions of around
$10^{-2}\,$sr, so that the yet detected sources already shadow more
than 10\% of the sky. 

Monte-Carlo simulations have shown that the significance of
detection for an isolated EGRET source for the energy selection
$E>$100 MeV is adequately represented as
\begin{equation}
\label{sig}
S=\alpha\, F_8 \sqrt{{E_7}\over {B_5}}
\end{equation}
where $F_8$ is the source flux in units of $10^{-8}\,\,{\rm cm^{-2}\,
sec^{-1}}$, $E_7$ is the exposure in units of $10^7\,\,{\rm cm^2\,sec}$,
$B_5$ is the intensity of background emission in the region of the source
in units of $10^{-5}\,\,{\rm cm^{-2}\,sec^{-1}\,sr^{-1}}$, and $\alpha=0.08$
is an empirically determined constant (\cite{mat96}). With the standard
thresholds of $5\sigma$ for $\vert b\vert \le 10^\circ$ and $4\sigma$ outside
the galactic plane we can use (\ref{sig}) to calculate the sensitivity threshold
$F_c$ for each viewing direction. To determine $E_7$ we have summed
the exposure of all viewing
periods of the Phases 1-4 corresponding to observations between April 1991
and September 1995. The background intensity is the sum of the diffuse
galactic emission (\cite{hunt97}), the extragalactic background
(\cite{sre97}), and the
point-spread functions of all $> 5 \sigma$ excesses which have been
found in a Maximum-Likelihood search (\cite{mat96}) in the summed data
of Phases 1-4.

The sensitivity threshold $F_c$ corresponds to the integrated
flux above 100 MeV. Thus the critical distance up to which a pulsar can
be detected directly has to be calculated as

\begin{equation}
X_{max} (l,b,t) =\sqrt{{\sum\limits_{k=3}^9\, S_m(t,E_k)\, \delta E_k}
\over {F_c (l,b)}} \qquad {\rm kpc}\ \end{equation}
where $\delta E_k$ is the width of the energy bin $E_k$.
It can be expected that out of the galactic plane $X_{max}$ generally
exceeds the line-of-sight through the galactic disk, so that all pulsars
would be detectable by EGRET. Thus only in the galactic plane pulsars can
contribute to the diffuse \gr emission.

\section{The contribution to the diffuse emission}

The mean velocity of nearby pulsars is $\sim 450\,
\,{\rm km\,sec^{-1}}$ (\cite{Lylo94}), and thus pulsars typically travel over
a distance of 1 kpc in
2 million years. This is much larger than the vertical scale height,
but considerably smaller than the radial scale length of the spatial 
distribution of pulsar birth locations (\cite{pac90}). Thus the vertical 
distribution of pulsars will strongly depend on their age, and the
radial distribution will remain basically unaffected. 
It is straightforward to show that the pulsar trajectories are not
strongly influenced by the gravitational potential of the Galaxy, except
for the innermost kiloparsec around the galactic center.
The potential gradients in vertical direction
change the mean pulsar velocity only at the percent level over a million years.
The radial gradients in the plane are balanced to first order
by the galactic rotation. 

We may thus parametrize the normalized spatial distribution of pulsar
birth locations in cylindrical coordinates ($r,z$) as

\begin{equation}
\rho_b (r,z) ={{0.0435\  \exp(-{{\mid z \mid}\over {z_c}})}\over {z_c\, 
r_c^2\ \cosh \left({{r}\over {r_c}}\right)}} 
\end{equation}
where the $\cosh$-term accounts for the galactocentric gradient.
The normalized spatial distribution of pulsars at time $t$ is then derived by
integration over the 3D distribution of pulsar birth velocities $P(v)$ 
(\cite{Lylo94})
\begin{equation}
\rho (r,z,t)= {{0.0218}\over {z_c\, 
r_c^2\ \cosh \left({{r}\over {r_c}}\right)}}
\int_{-1}^1 dx \int_0^\infty dv\ P(v)\, \exp\left(- {{\mid z+vxt \mid}
\over {z_c}} \right)
\end{equation}

With $\tau_p^{-1}$ as the birth rate of
\gr pulsars and $\epsilon $ as the fraction which radiates in
our direction we can determine the number of directly detectable pulsars 
by integration over the line-of-sight, solid angle, and age
\begin{equation}
N_{det} = {{\epsilon}\over {\tau_p}} \int_0^{t_{max}} dt\ 
\oint d\Omega\ \int_0^{X_{max}} dx\ x^2\,\rho (r,z,t)
\end{equation}
as well as the diffuse emission of unresolved pulsars
\begin{equation}
\label{diffemis}
I_{dif} (E_k) = {{\epsilon}\over {\tau_p\,\Omega}} \int_0^{t_{max}} dt\ 
\int_{\Omega} d\Omega^\prime\ \int_{X_{max}}^\infty dx\ \rho (r,z,t)\, 
S_m(t,E_k) 
\end{equation}
where

\begin{equation}
z=x\,\sin b\ ,
\quad r=\sqrt{r_\odot^2 +x^2 \cos^2 b - 2\,r_\odot x \cos b \cos l}
\end{equation}
Please note that any change in the total number of pulsars has similar
impact on the number of directly observable pulsars as on the \gr intensity
of unresolved objects. Thus the direct detections provide a strong constraint
on the pulsar contribution to the diffuse galactic \gr emission and limit
our choice of $\epsilon/\tau_p$.

\begin{figure}
\plotone{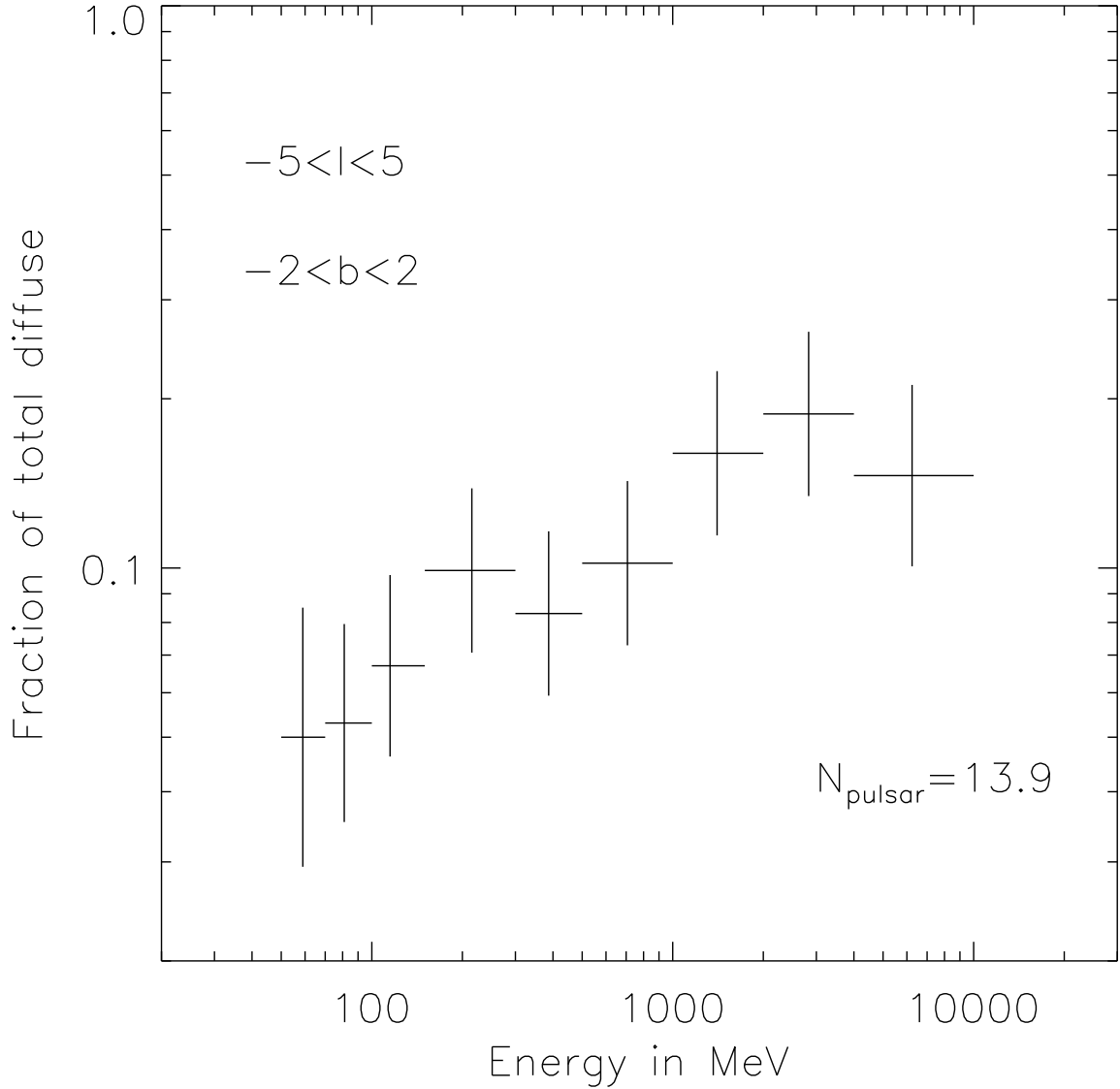}
\caption{Here we show the fraction of the total diffuse intensity in direction
of the Galactic Center (Hunter et al. 1997)
which can be attributed to unresolved 
pulsars. The parameters to the model are given in the text. 
The error bars are derived by propagation of the parameter uncertainty
in the pulsar model fit (see Table 1).
The intensity due to pulsars scales almost linearly with the number of objects
which are supposed to be seen as point sources, in this case 
13.9 pulsars. \label{spec0}}
\end{figure}

\placefigure{spec0}

\begin{figure}
\plotone{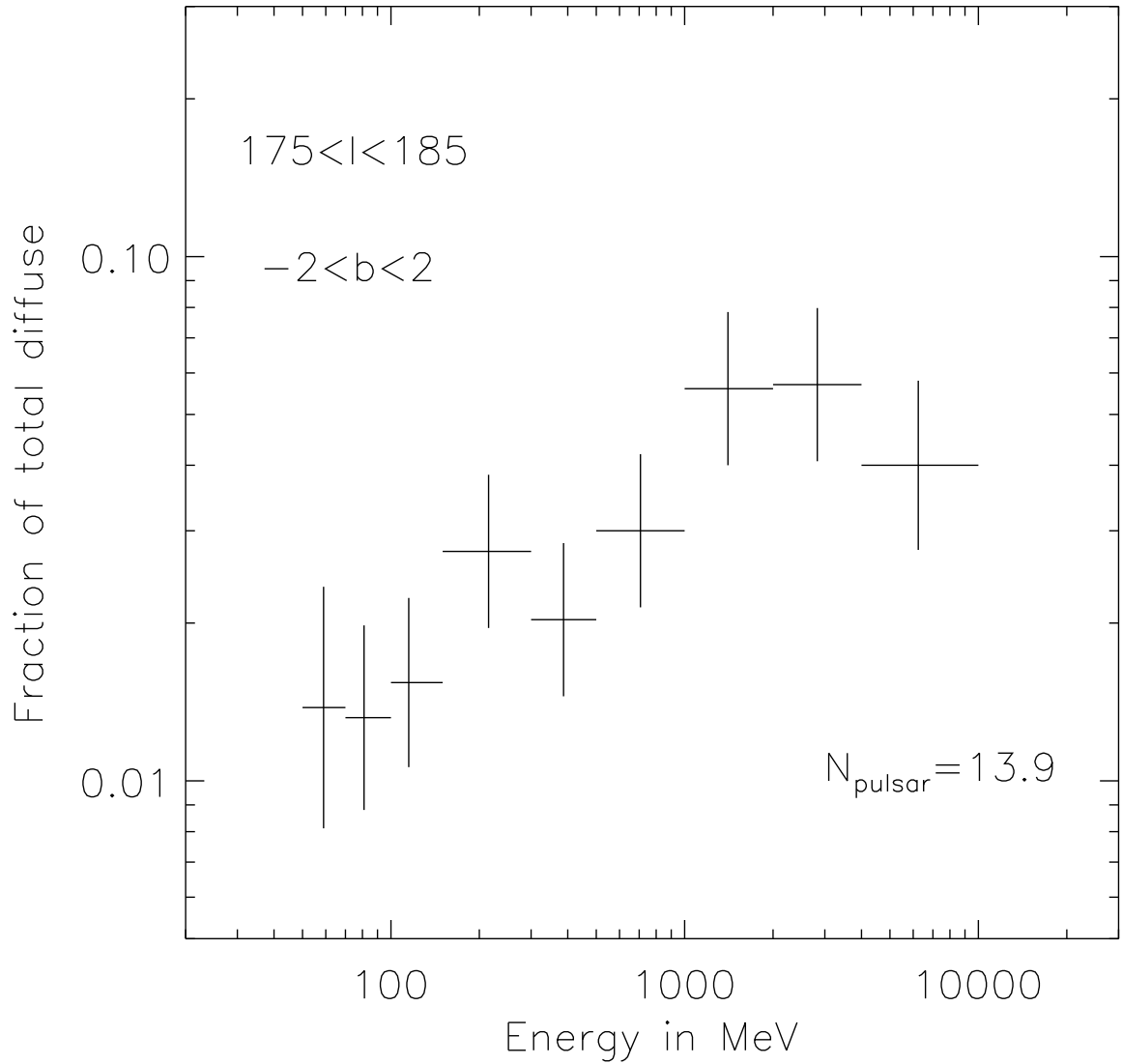}
\caption{Here we show the fraction of the total diffuse intensity in direction
of the anticenter which can be attributed to unresolved 
pulsars. Again the parameters to the model are given in the text.  
\label{spec180}}
\end{figure}

\placefigure{spec180}

With (\ref{diffemis}) we can now calculate the spectrum of unresolved pulsars and
compare it to the spectrum of the total observed diffuse emission. This is
shown for the Galactic Center direction in Fig. \ref{spec0} and for the anticenter in
Fig. \ref{spec180}, where we plot the fraction of the total observed emission which is 
due to pulsars. The data have been deconvolved to correct for the
effects of EGRET's point-spread-function (\cite{hunt97}).
A radial scale length $r_c$=3.5 kpc
was used for the spatial distribution of pulsars.
The maximum age $t_{max}$ used is $1.6\cdot 10^6\,$years, though the spectra depend only weakly on age;
the age affects the number of 
directly observable objects $N_{det}$ nearly as much as the predicted 
emission of unresolved pulsars $I_{dif}$. In addition, a larger $t_{max}$
would require extrapolation too far beyond the range of pulsar ages for which we
have data. As we will see later, $t_{max}$ does
have a strong influence on the latitude distributions of observable pulsars
as well as of the pulsar contribution to the galactic diffuse emission.

The parameters used here ($\tau_p^{-1} = 0.01\ {\rm years^{-1}}$
and $\epsilon = 0.15$) imply that around 14 pulsars should be detectable
by EGRET, which is to be compared to a total galactic population of
$t_{max} \tau_p^{-1} \epsilon = 2400$ pulsars which radiate in our direction. 
Since six are already identified this would mean that another
eight unidentified sources are actually pulsars. 
Considering the integrated
intensity above 100 MeV in this example pulsars would cause 10\% of the 
observed intensity in the Galactic Center direction and about 3\% in the
anticenter direction. Pulsars contribute very little off the plane.
In direction l$\simeq$0$^{\circ}$, b$\simeq$8$^{\circ}$ we get around 1\%,
so that integrated over the sky pulsars provide only a few percent of the
observed diffuse emission, consistent with earlier estimates 
based on pulsar emission models. This is also true for the emission at higher
\gr energies.
We have compared the latitude distribution of the
observed diffuse emission above 1 GeV
to what our model predicts for the pulsar contribution.
The result is given in Fig. 4 where we show the distribution of the diffuse
intensity at energies above 1 GeV as averages over longitude.
We find that pulsars contribute strongly only very close to the plane where
the lines-of-sight are long, but the observed emission does fall off
much less rapidly with latitude than the emission of unresolved pulsars.
Also the observed spectral discrepancy above 1 GeV seems
to extend to latitudes $|b|\ga 5^{\circ}$ (\cite{hunt97}).

If we would allow for higher pulsar ages, say $t_{max}= 5\cdot 10^6$ years,
the latitude distribution of the emission of unresolved pulsars would
get close to that of the observed emission. However, for the same number of
directly observable pulsars only around 6\% of the observed diffuse emission
above 1 GeV would be caused by unresolved pulsars, as shown by the dotted
line in Fig. 4. In our study the beaming 
fraction $\epsilon$ is taken to be a constant. Any physical variation of
$\epsilon$ in the regime of observed pulsar ages ($10^3 - 10^{5.7}$ years)
should be accounted for by our fit parameters $y_k$ and $b_k$. Any 
variations of $\epsilon$ at higher pulsar ages cannot be accounted for.
The fact that no pulsar with rotational age of 1 million years or higher
has been observed indicates strongly that \gr emission of pulsars ceases 
at ages of roughly 1 million years. Note that for $t_{max} =5\cdot 10^6$
years more than 50\% of all observable pulsars in our model
would be older than 1 million years. Hence the results 
we present for a high $t_{max}$ serve to demonstrate the basic behaviour
rather than to provide a realistic estimate.

Many of the observable pulsars should be located at
higher latitudes. We expect 58\% of pulsars at $|b|\ga 6^{\circ}$ (74\%
in case of $t_{max} =5\cdot 10^6$ years) and 31\% at latitudes
$|b|\ga 14^{\circ}$ (53\% in case of $t_{max} =5\cdot 10^6$ years).

\begin{figure}
\plotone{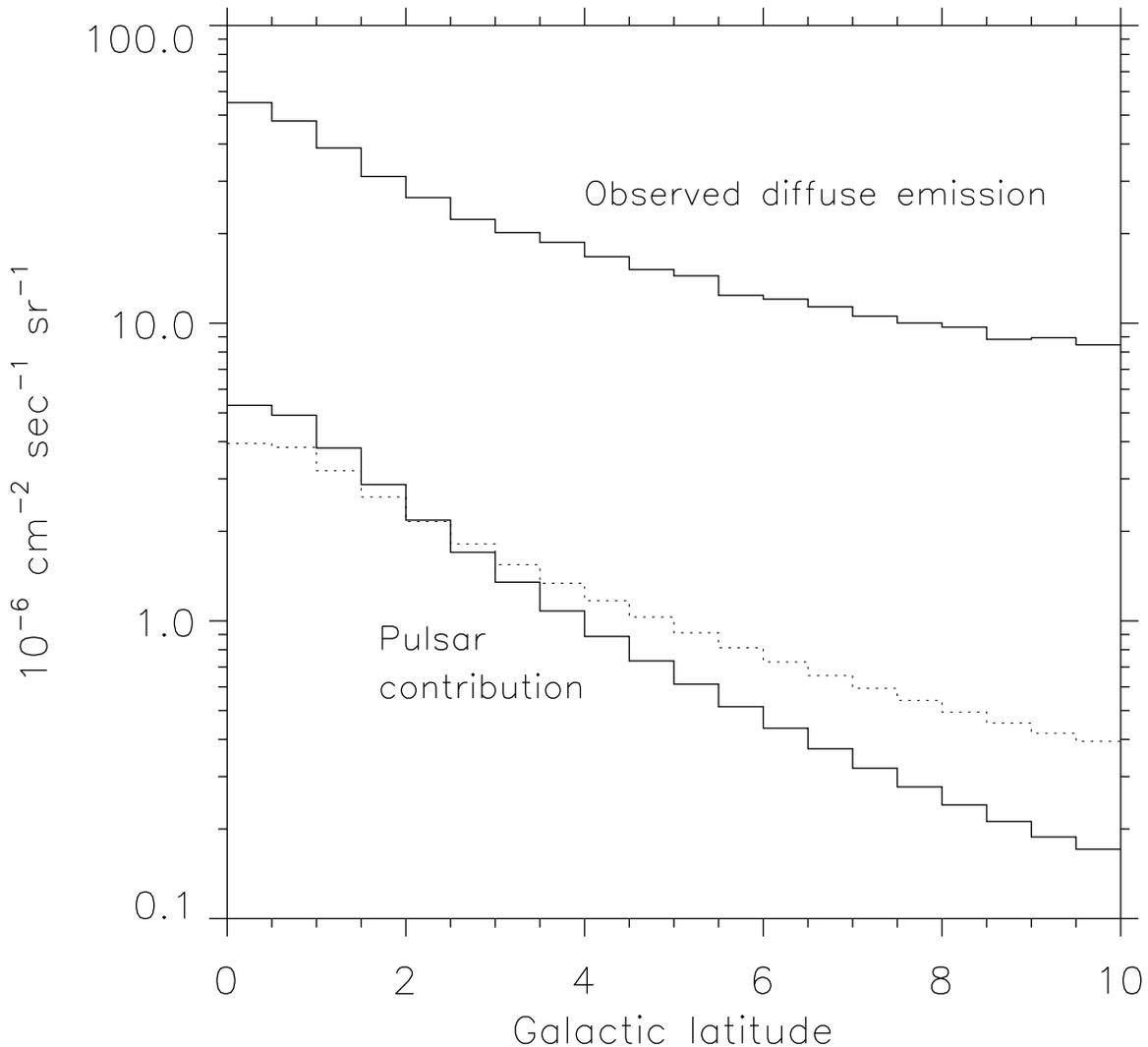}
\caption{The latitude distribution of the diffuse emission
above 1 GeV as observed by EGRET (Hunter et al. 1997)
is much wider than that of the pulsar
contribution, here shown as solid line for 13.9 directly observable pulsars.
Both the data and the model have been averaged over longitude. In
contrast to Fig.2 and Fig.3 we show the original observed intensity
distribution compared to the psf-convolved model prediction. The statistical 
error of the observed emission is at the percent level and thus negligible.
The dotted histogram shows what we would get if the maximum age $t_{max}$ were
5 million years. The latitude distribution gets closer to that of the
observed diffuse emission, but after renormalization to 13.9 observable
sources (by a factor 0.56) the pulsar contribution to the diffuse emission is 
only around 6\%.
\label{lat}}
\end{figure}

\placefigure{lat}

\section{Discussion}


The dominant contribution of unresolved pulsars to the diffuse \gr emission is above 1 GeV, where in
direction of the Galactic Center around 18\% of the total observed emission
could be provided with approximately 14 directly observable objects. 
Approximately 30 pulsars would have to be directly observable to account for 100\% of the discrepancy
between model predictions and the observed spectrum in this direction.
This seems unreasonable since only few of the
unidentified EGRET sources show the typical pulsar-like \gr spectrum
(\cite{merc96}). Furthermore, studies of known radio pulsars have led
to constraining upper limits for pulsed emission for most of them
(\cite{thom94}, \cite{nel96}). 

A few pulsars are positionally coincident with
unidentified EGRET sources but it is unclear whether the \gr emission is
caused by the pulsar, the supernova remnant, or any other nearby system.
A deep search for radio pulsars in the error boxes of 19 unidentified 
EGRET sources has been performed recently (\cite{ns97}). Though data were
taken at three different frequencies to minimize selection effects, no new
pulsars were found. 
So the detectable but not yet detected pulsars would have to be
either radio-quiet or located in active regions like SNOB's 
(\cite{mont79}) which would impede the radio detection by
their high and possibly variable dispersion measures. It is interesting to see
that roughly 10 unidentified EGRET sources can be associated either with
SNR or with OB associations or with both 
(\cite{stur95}, \cite{kaar96}, \cite{yadi96}). However, supernova remnants only live
for around $10^5\,$years, so that preferentially young pulsars would be
expected to be associated with them, whereas the bulk of the observable
pulsars should be old. From our modeling we expect 24\% of the observable pulsars at an age less than $10^5\,$years and 76\% older than this
(for $t_{max} =1.6\cdot 10^6\ $years).
Therefore we should not assume too many young pulsars hidden in regions of high 
electron density without having a corresponding number of old pulsars.
It thus seems reasonable that a only a few \gr pulsars are hidden
radio pulsars.

It is also possible that there is a substantial number of radio-quiet
but \gr loud pulsars. Though an extensive search for \gr pulsars among
the brightest unidentified EGRET point sources has obtained only upper
limits to date (\cite{matt96}),
the spatial and flux distribution of unidentified EGRET sources has been
reported to be compatible with the majority of them being
pulsars (\cite{yadi96}). If these were all Geminga-like, the corresponding
unresolved sources would contribute strongly to the diffuse emission above
1 GeV. Since our analysis is based on the observed properties of mainly
radio-loud pulsars, any basic difference in the \gr emission properties
between radio-loud and radio-quiet objects would severely affect the
predictions and leave us without a reliable tool to estimate numbers.
This would be true even if the recent detection of weak pulsed radio
emission from Geminga can be confirmed (\cite{kl97}).

Nevertheless, the small fraction of unidentified EGRET sources which
shows pulsar-like or Geminga-like \gr spectra (\cite{merc96})
argues strongly against
the bulk of unidentified sources being pulsars. Also a substantial
fraction of the unidentified EGRET sources in the galactic plane
appears to be variable which makes the identification of {\it all}
unidentified sources with pulsars even more unlikely (\cite{mcla96}).

\section{Conclusion}
In total, we think that around 14 directly observable pulsars, of which six
are already identified, is a reasonable number, and that thus the calculated
contribution of unresolved pulsars can be taken as serious estimate. 
The main systematical uncertainty in our study definitely is that the six
identified pulsars may not be a representative sample. On the other hand,
we do not need to base our study on a theoretical model which may or may
not be a fair description of nature.

We find that pulsars contribute very little to the diffuse emission at lower
energies, whereas above 1 GeV they can account for 18\%
of the observed intensity in selected regions.
This seems not to be enough to explain the discrepancy between observed  
intensity and model predictions based on cosmic-ray interactions in this
regime. We also find that pulsars contribute mainly close to the plane where
the lines-of-sight are long, but the observed emission does fall off
much less rapidly than this.
Also the observed spectral discrepancy between the data above 1 GeV
and the predictions of cosmic ray models seems
to extend to latitudes $|b|\ga 5^{\circ}$ (\cite{hunt97}). Thus there must be
additional effects playing a r\^ ole for the observed diffuse \gr emission
above 1 GeV. 

\acknowledgements

The EGRET Team gratefully acknowledges support from the following:
Bundesministerium f\"{u}r Bildung, Wissenschaft, Forschung und Technologie
(BMBF), Grant 50 QV 9095 (MPE); NASA Cooperative Agreement NCC 5-93 (HSC); 
NASA Cooperative Agreement NCC 5-95 (SU); and NASA Contract NAS5-96051
(NGC).

\clearpage
 
\begin{deluxetable}{cccrcc}
\footnotesize
\tablecaption{Results of the fit of the ideal pulsar to
the data of the six observed objects. The energy range in
column 1 is given in units of GeV. The parameters $b$ and $y$
determine the model in Eq.1. The $\chi^2$ sums in column 6 are around four
which is the number of degrees of freedom. \label{tbl-1}}
\tablewidth{0pt}
\tablehead{
\colhead{Energy} & \colhead{b} & \colhead{$\delta$b}   & \colhead{y} & 
\colhead{$\delta$y}  & \colhead{$\chi_{min}^2$}}
\startdata
0.05-0.07 & 0.94 & 0.19 & -7.40 & 0.16 & 3.78 \nl
0.07-0.10 & 0.89 & 0.20 & -7.63 & 0.15 & 5.50 \nl
0.10-0.15 & 0.80 & 0.18 & -7.88 & 0.15 & 3.38 \nl
0.15-0.30 & 0.61 & 0.14 & -8.24 & 0.14 & 5.35 \nl
0.30-0.50 & 0.64 & 0.14 & -8.71 & 0.14 & 3.36 \nl
0.50-1.00 & 0.57 & 0.13 & -9.11 & 0.14 & 3.64 \nl
1.00-2.00 & 0.39 & 0.13 & -9.72 & 0.13 & 5.05 \nl
2.00-4.00 & 0.43 & 0.14 & -10.27 & 0.14 & 2.32 \nl
4.00-10.0 & 0.44 & 0.15 & -11.25 & 0.15 & 3.60 \nl

\enddata
 
\end{deluxetable}

\clearpage

\clearpage

\clearpage

\clearpage

\end{document}